\documentstyle[aps,prl,multicol,epsf]{revtex}

\begin{document}
\widetext
\title{\bf The boson-fermion model with on-site Coulomb repulsion
between fermions}

\author{Alfonso Romano}

\address{Dipartimento di Fisica ``E.R. Caianiello",
Universit\`a di Salerno, I-84081 Baronissi (Salerno), Italy\\
Unit\`a I.N.F.M. di Salerno}

\date{\today}
\maketitle \draft
\begin{abstract}
The boson-fermion model, describing a mixture of itinerant
electrons hybridizing with tightly bound electron pairs
represented as hard-core bosons, is here generalized with the
inclusion of a term describing on-site Coulomb repulsion between
fermions with opposite spins. Within the general framework of the
Dynamical Mean-Field Theory, it is shown that around the symmetric
limit of the model this interaction strongly competes with the
local boson-fermion exchange mechanism, smoothly driving the
system from a pseudogap phase with poor conducting properties to a
metallic regime characterized by a substantial reduction of the
fermionic density. On the other hand, if one starts from
correlated fermions described in terms of the one-band Hubbard
model, the introduction in the half-filled insulating phase of a
coupling with hard-core bosons leads to the disappearance of the
correlation gap, with a consequent smooth crossover to a metallic
state.

\end{abstract}

\pacs{PACS numbers: 71.10.Fd, 71.30.+h, 71.10.Fd, 74.20.Mn }

\begin{multicols}{2}

\narrowtext

A large variety of phenomena in condensed matter physics has been
studied in the past in terms of interacting models with coupled
bosonic and fermionic degrees of freedom \cite{Mahan}. Their
relevance has proved to be of particular evidence in the analysis
of the electron-phonon problem and of the related issue of the
polaron and bipolaron formation. In this context, it was proposed
some years ago \cite{Ran85} that for a system of itinerant
electrons interacting with local lattice deformations, the
crossover regime between adiabatic and non-adiabatic behavior can
be described by a model where tightly bound electron pairs
(hard-core bosons) of polaronic origin coexist with quasifree
electrons (fermions), with an exchange coupling assumed between
them by which bosons can decay into pairs of itinerant fermions
and vice-versa. It has subsequently been suggested \cite{bfm} that
this Boson-Fermion Model (BFM) can provide a possible scenario for
high-$T_c$ superconductivity, according to the hypothesis that the
fermionic degrees of freedom describe holes confined in the
copper-oxygen planes, while the bosonic ones are associated with
bipolarons which form in the highly polarizable dielectric layers
sandwiching the CuO$_2$ planes. Since then, the model has been
deeply investigated, in particular by Ranninger and coworkers
\cite{pseu,RRR,hallbfm}, with a special attention to the
description of the pseudogap phase characterizing the normal state
of the underdoped high-$T_c$ copper oxides.

In this paper we consider an extension of the BFM in which
fermions, besides exchanging with bosons, are assumed to also
experience an on-site Coulomb repulsion. This interaction, tending
to prevent two fermions with opposite spin from occupying the same
site, strongly competes with the boson-fermion exchange mechanism,
giving rise to interesting effects in the spectral as well as in
the transport properties.  The model, whose fermionic part goes
back in the one-band Hubbard model in the limit of vanishing
boson-fermion coupling, is solved here within the framework of the
Dynamical Mean-Field Theory (DMFT)
\cite{Geo92,Jar92,Jar93,DMFT,Geb}, by applying to the related
self-consistent impurity problem the so-called Non-Crossing
Approximation (NCA) \cite{NCA}. We recall that DMFT-based
approaches have been intensively used in the last years in the
related context of electron-phonon systems described by the
Holstein model \cite{elphdmft,Ciuchi} or extensions of it
\cite{Millis}, as well as in the case of boson-fermion coupled
systems with boson dispersion \cite{Kot}. By means of various
analytical and numerical techniques, a special attention has also
been devoted to the role played by the on-site Coulomb repulsion
in the electron-phonon problem and, in particular, on the polaron
and bipolaron formation. This has been done essentially referring
to a generalized version of the Holstein model which includes an
Hubbard interaction term (the so called Holstein-Hubbard model)
\cite{HH}.

The hamiltonian for the BFM with Coulomb repulsion between
fermions is given by
\begin{eqnarray}
H & = &
\varepsilon_0\sum_{i,\sigma}c^{\dagger}_{i\sigma}c_{i\sigma}
-t\sum_{\langle i j\rangle,\sigma}c^{\dagger}_{i\sigma}c_{j\sigma}
+U\sum_i c^{\dagger}_{i\uparrow}c_{i\uparrow}
 c^{\dagger}_{i\downarrow}c_{i\downarrow}
\nonumber \\ & & + \; E_0 \sum_i b^{\dagger}_i b_i + g \sum_i
[b^{\dagger}_i c_{i\downarrow} c_{i\uparrow} +
c^{\dagger}_{i\uparrow} c^{\dagger}_{i\downarrow} b_i] \quad .
\label{ham}
\end{eqnarray}
\noindent Here $c_{i\sigma}^{(\dagger)}$ denote  annihilation
(creation) operators for electrons with spin $\sigma$ at site $i$
and $b_i^{(\dagger)}$ denote hard-core bosonic operators
describing tightly bound localized electron pairs. The site
energies for fermions and bosons are expressed as
$\varepsilon_0=D-\mu$ and $E_0=\Delta_B-2\mu$, respectively, where
the chemical potential $\mu$ is assumed to be common to the two
kinds of carriers (up to a factor 2 for the bosons) in order to
guarantee charge conservation. Finally, $g$ denotes the
boson-fermion pair-exchange coupling constant, while $t$, $D$, and
$U$ are respectively the bare hopping integral, the bare half
bandwidth and the on-site Coulomb repulsion, all referred to
fermions. We see that in the limit of vanishing Coulomb repulsion
the model goes back in the boson-fermion model for high-$T_c$
superconductors analyzed in Refs.\cite{bfm,pseu,RRR,hallbfm}. On
the other hand, in the limit of zero boson-fermion coupling the
fermionic part of the model defined by the Hamiltonian (\ref{ham})
coincides with the standard one-band Hubbard model.

It is worth pointing out that the assumption of an on-site Coulomb
interaction affecting only the fermionic degrees of freedom is
motivated by the existence of anisotropic systems, such as the
previously mentioned high-$T_c$ copper oxides, where there is
experimental evidence that bosons and fermions reside in separated
lattice regions where Coulomb effects are in general significantly
different. In this respect, the site indices in the
phenomenological Hamiltonian (1) would better be thought as
representative of molecular units composed, for instance, by an
atom on which the correlated fermions sit and two adjacent ones
forming a diatomic molecule housing the bosonic tightly bound
electron pairs. Given this picture, no direct interference occurs
between the on-site fermionic Coulomb repulsion and the bosonic
formation, the only effect being a shift at higher energies of the
fermionic level with respect to the bosonic one, leading to a
relative change in occupation of the two species. The assumption
of a lattice seen as a collection of effective sites is of course
made to simplify the analytical as well as the numerical
calculations, and a more accurate (though technically more
complicate) way of analyzing the role of the Coulomb repulsion
would require the solution of the model on a true bipartite
lattice.

Given the local character of the interaction terms in the
Hamiltonian (\ref{ham}), important insights on the physics of the
model can be obtained from its atomic limit. The eigenstates and
eigenvalues of $H$ are in this case given by
\begin{equation}
\begin{array}{ll}
| 1 \rangle \; = \; | 0 \rangle & E_{1} \; = \; 0 \\ | 2 \rangle
\; = \; | \uparrow \rangle & E_{2} \; = \; \varepsilon_{0}
\\ | 3 \rangle \; = \; | \downarrow \rangle & E_{3} \; = \;
\varepsilon_{0} \\ |4 \rangle \; = \; u | \uparrow \downarrow
\rangle \; - \; v | \bullet \rangle \qquad & E_{4} \; = \;
\varepsilon_{0} + U/2 + E_{0}/2 - \gamma \\ | 5 \rangle \; = \; v
| \uparrow \downarrow \rangle \; + \; u | \bullet \rangle & E_{5}
\; = \; \varepsilon_{0} + U/2 + E_{0}/2 + \gamma \\ | 6 \rangle \;
= \; | \uparrow  \; \bullet \rangle & E_{6} \; = \;
\varepsilon_{0} + E_{0} \\ | 7 \rangle \; = \; | \downarrow  \;
\bullet \rangle & E_{7} \; = \; \varepsilon_{0} + E_{0} \\ | 8
\rangle \; = \; | \uparrow \downarrow \bullet \rangle & E_{8} \; =
\; 2 \varepsilon_{0} + U + E_{0}
\end{array}
\label{eigen}
\end{equation}
where the dot indicates the presence of a boson and we have
defined
\begin{eqnarray}
u^{2} \; &=& \; \frac{1}{2} \left[ \; 1 \; - \;
\frac{2\varepsilon_{0}+U - E_{0}}{2\gamma} \; \right] \nonumber \\
\gamma \; &=& \; {1\over 2}\, \left[ \; (2\varepsilon_{0}+U -
E_{0})^{2} \; + 4\; g^{2} \; \right]^{1/2} \quad
\end{eqnarray}
with $u^2+v^2=1$. When $U=0$ and the total particle density
$n=n_F+2n_B$ is equal (or close) to 2, the physics underlying the
model is fully determined by the charge exchange interaction. As
the temperature is decreased, the spectral weight of the pole in
the Green's function associated with the non-bonding
single-particle state (at energy $\varepsilon_0$) is gradually
transferred to the poles (at energies $E_4-\varepsilon_0$ and
$E_5-\varepsilon_0$) associated with the bonding and anti-bonding
two-particle states $|4>$ and $|5>$ \cite{Dom98}. The
corresponding effect seen when itinerancy is taken into account,
is a depletion of the density of states in the region around the
chemical potential. This gives rise to the opening a pseudogap
which, according to the mechanism explained above, is thus induced
by local pairing among the electrons.

When the effect of the on-site Coulomb repulsion among fermions is
introduced, the structure of the atomic Green's function becomes
more complex, with the number of poles increasing from 3 to 6. The
transfer of spectral weight from single-particle to two-particle
resonances as $T$ is decreased, becomes less and less pronounced
as higher values of $U$ are considered. As expected, this is a
consequence of the fact that a finite $U$ tends to forbid double
fermion occupations of the same site, thus inhibiting the
boson-fermion charge exchange mechanism. More precisely, what one
can see from the atomic limit solution is that, upon increasing
$U$, $u^2$ tends to zero, $v^2$ tends to 1, and thus the two
states $|4>$ and $|5>$ tend to become pure one-boson and
two-fermion states, respectively, with a separation in energy of
the order of $U$. As we will see in the following, for the full
model around the symmetric limit this has the simultaneous effect
of hindering the opening of the pseudogap and making the bosonic
site occupation considerably higher than the fermionic one.

The effect of the electron itinerancy will now be taken into
account within the general framework of the Dynamical Mean-Field
Theory (DMFT) \cite{Geo92,Jar92,Jar93,DMFT,Geb}. In the last years
this approach has proved to be of fundamental importance in
condensed matter physics and has been successfully applied to a
variety of many-body models. Within DMFT an interacting system is
seen as a purely local system coupled to a Weiss ``mean field'' to
be determined self-consistently, produced by the neighboring
sites. This approach leads to the freezing of all spatial
fluctuations, just like the usual classical mean-field theory, but
has the advantage of fully retaining the quantum temporal
fluctuations of the original problem (hence the adjective
``dynamical''). Indeed, the mean field is here a function of time,
instead of being a pure number, associated with the probability
amplitude of creating a particle on the impurity center at a given
time and destroying back in the external bath at a later time, and
vice-versa. Like any mean field approximation, this approach
becomes more and more accurate as the lattice coordination number
increases, becoming exact in the limit of infinite dimensionality.
This basic structure makes the application of the DMFT
particularly suited to models, such as that defined by the
Hamiltonian (\ref{ham}), where the interaction terms are local. A
method for the inclusion of non-local corrections to DMFT, known
as Dynamical Cluster Approximation, has recently been developed
\cite{DCA} and applied to the two-dimensional Hubbard model
\cite{Jar00}, by mapping the lattice problem onto a periodic
cluster embedded in an external host.

Since we consider here a model where correlated electrons are
coupled to {\it dispersionless} bosons, in the DMFT
self-consistent procedure the bosonic channel for the Green's
function remains frozen, in the sense that the related bare Weiss
self-energy at each iteration is unrenormalized and fixed at its
initial non-interacting expression. This point, discussed at
length in Ref.\cite{Kot}, implies that the general dynamical
mean-field framework in which the model defined by the Hamiltonian
(\ref{ham}) is investigated, is essentially the same as that
developed for the Hubbard model in Refs.\cite{Geo92,Jar92,Jar93},
the effect of the boson-fermion exchange coupling manifesting
itself only in the solution of the impurity problem. In the
following we give a brief summary of the DMFT technique, along the
general lines of the diagramatic approach presented in
Refs.\cite{Jar92,Jar93}. The reader is referred to these papers,
as well as to Refs.\cite{Geo92,DMFT,Geb}, for all technical
details not reported here for brevity.

The approach starts from the demonstration \cite{Metz,Mul89} that
in the limit of infinite dimensionality only the local dynamics
remains nontrivial, in the sense that the only non-vanishing
contributions to the self-energy are the site-diagonal ones. As a
consequence, the Dyson equation in real space takes the form
\begin{equation}
G_{ij}(i\omega_n) = G_{ij}^0(i\omega_n) + \sum_l G_{il}(i\omega_n)
\Sigma(i\omega_n) G_{lj}(i\omega_n) \quad .
\end{equation}
When this equation is referred to the local propagator $G_{ii}$,
one can first consider the self-energy contributions coming from
all sites $j$ other than the site $i$. The resummation of these
contributions lead to the definition of a modified propagator,
that we denote by $G_W$ (the Weiss field), that is then used in
the full Dyson equation for $G_{ii}$, reintroducing the missing
self-energy contributions that involve the same site $i$
\cite{Jar93}. This procedure leads to the following Dyson equation
for $G_{ii}\equiv G_{imp}$
\begin{equation}
G_{imp}(i\omega_n)=\left[G_W^{-1}(i\omega_n)-\Sigma(i\omega_n)\right]
\label{gimpdy} \quad ,
\end{equation}
which proves the equivalence of the lattice problem to a model of
an Anderson impurity self-consistently embedded in a medium
specified by the Weiss Green's function $G_W$. Given for the
latter the structure
$G_W(i\omega_n)=[i\omega_n-\varepsilon_0-\Sigma_W(i\omega_n)]^{-1}$,
the above equation implies that the impurity Green's function can
be expressed in the form
\begin{equation}
G_{imp} (i\omega_n)= {1 \over i \omega_n - \varepsilon_0 -
\Sigma_W(i\omega_n) - \Sigma_{int}(i\omega_n)} \quad ,
\label{gimp}
\end{equation}
where we have put $\Sigma_{int}\equiv\Sigma$. We thus see that the
total self-energy is written as the sum of two
momentum-independent contributions $\Sigma_{int}$ and $\Sigma_W$,
associated respectively with the on-site interactions (the
boson-fermion exchange and the Coulomb repulsion among fermions)
and with the hybridization of the impurity center with the medium.

On the other hand, due to the locality of the self-energy, the
lattice Green's function in ${\bf k}$-space is given by
\begin{eqnarray}
G_{lat}(\varepsilon_{\bf k},i\omega_n) = {1 \over i \omega_n -
\varepsilon_0 -\varepsilon_{\bf k} - \Sigma_{int}(i\omega_n)}
\label{glat}
\end{eqnarray}
where $\varepsilon_{\bf k}$ denotes the bare electron dispersion.
The self-consistency condition which iteratively leads to the
renormalization of the Weiss self-energy, thus closing the DMFT
algorythm, is then simply obtained by equating the impurity
Green's function (\ref{gimp}) to the integral over ${\bf k}$-space
of the lattice Green's function, i.e.
$G_{imp}(i\omega_n)=N^{-1}\sum_{\bf k}G_{lat}(\varepsilon_{\bf
k},i\omega_n)$. Upon replacement of the integration over ${\bf k}$
by an integration over energy, this condition takes the form
\begin{equation}
G_{imp}(i\omega_n)= \int d\varepsilon \,{\rho(\varepsilon) \over
i\omega_n - \varepsilon_0 - \varepsilon - \Sigma_{int}(i\omega_n)}
\quad , \label{self}
\end{equation}
where $\rho$ is the free density of states associated with the
particular lattice chosen.

The whole DMFT procedure can thus be summarized as follows. One
starts with a guess for the Weiss Green's function $G_W$ and solve
the corresponding impurity problem, obtaining $G_{imp}$ (here this
is done by means of the so-called Non-Crossing Approximation). The
resulting local self-energy $\Sigma_{int}\equiv\Sigma$, evaluated
from Eq.(\ref{gimpdy}), is then used to find a new impurity
Green's function by imposing the self-consistency condition
(\ref{self}). By subtracting off from the latter the local
self-energy, again by use of Eq.(\ref{gimpdy}), one gets a new
Weiss field which is used to reinitialize the process, which is
then iterated until convergence is reached.

Let us also notice that the explicit form taken by the
self-consistency condition depends on the particular lattice on
which the DMFT equations are solved. In the following we assume to
refer to a Bethe lattice with coordination number $z\to\infty$ and
nearest neighbor hopping $t_{ij}=t/\sqrt{z}$, for which the
electronic bare density of states is the semi-circular function
$\rho(\varepsilon)=(1/2\pi t^2)\sqrt{4t^2-\varepsilon^2}$
($|\varepsilon|<2t$) of width $2D=4t$. This assumption, widely
used in the context of the DMFT
\cite{Ciuchi,Millis,Kot,Schork,Schork99} allows to solve
analytically eq.(\ref{self}), leading to
\begin{equation}
\Sigma_W(i\omega_n) \; = \; t^2 G_{imp}(i\omega_n)
\end{equation}
or, in terms of the corresponding imaginary parts,
\begin{equation}
\Delta(\omega)\;=\;t^2\;A_F(\omega)  \label{del}
\end{equation}
(we use the definition $\Delta(\omega)=-2\,{\rm
Im}\,\Sigma_W(\omega+i\eta)$, with a similar equation relating
$A_F$ and $G_{imp}$).

The single-site effective dynamics described above is naturally
investigated by means of functional methods based on the use of an
imaginary-time effective action for the fermionic degrees of
freedom defined at the impurity center \cite{Geo92,DMFT}. An
equivalent picture can be introduced by considering a related
Hamiltonian formulation where the Weiss field is described in
terms of auxiliary degrees of freedom. In the case of the BFM, we
have an effective Hamiltonian of the form
\begin{eqnarray}
\widetilde{H} & = & \sum_{\sigma} \varepsilon_{0}
c_{\sigma}^{\dagger} c_{\sigma} \; + \; U c_{\uparrow}^{\dagger}
c_{\uparrow}c_{\downarrow}^{\dagger} c_{\downarrow} \nonumber
\\ & & +\; E_{0} b^{\dagger} b \; + \; g \; [ \; c_{\uparrow}^{\dagger}
c_{\downarrow}^{\dagger} b \; + \; b^{\dagger} c_{\downarrow}
c_{\uparrow}]  \nonumber \\ & & + \; \sum_{k,\sigma} w_{k}
d_{k\sigma}^{\dagger}d_{k\sigma} \; + \; \sum_{k,\sigma} \; v_{k}
\; [ \; d_{k\sigma}^{\dagger} c_{\sigma} \; + \;
c_{\sigma}^{\dagger} d_{k\sigma} \; ] \; , \label{hamimp}
\end{eqnarray}
where $c_{\sigma}^{(\dag)}$ and $b^{(\dag)}$ denote the original
fermion and boson operators, respectively, at the selected site
and $d_{k\sigma}^{(\dag)}$ are the auxiliary fermionic operators
associated with the Weiss field, taking into account the dynamics
occurring at all the other sites. Their energy spectrum $w_{k}$
and their coupling $v_{k}$ to the impurity electrons are of course
not known {\it a priori} but must be determined self-consistently.
Identifying the Weiss Green's function with the propagator for the
auxiliary variables, one gets
\begin{equation}
G_W^{-1}(i\omega_n) = i\omega_n - \varepsilon_0 - \sum_k
{v_k^2\over{i\omega_n - w_k}} \quad ,
\end{equation}
with the sum at the right-hand side representing the self-energy
$\Sigma_W$. We thus see that the Hamiltonian description is fully
equivalent to the functional one provided that the parameters
$v_{k}$ and $w_{k}$ are related by the following equation to the
imaginary part $\Delta(\omega)$ of the Weiss self-energy appearing
in the effective action:
\begin{equation}
\Delta(\omega) \; = \; 2 \pi \; \sum_{k} \; v_{k}^{2} \;
\delta(\omega - w_{k}) \quad.
\end{equation}
Depending on the technique adopted to solve the impurity problem,
the Hamiltonian formulation can prove to be useful in practical
calculations, besided providing a description probably closer to
our physical intuition.

The self-consistent single-impurity Anderson problem defined by
the effective Hamiltonian (\ref{hamimp}) is solved here within the
so-called Non Crossing Approximation (NCA) \cite{NCA}, along the
lines presented and discussed in Ref.\cite{RRR}. Within the DMFT
framework, NCA has been applied in the last years to several other
models, such as the one-band Hubbard model \cite{Pru93}, a
multiband Hubbard model for perovskites \cite{Avignon}, the
Anderson lattice model with correlated conduction electrons
\cite{Schork}, and the Kondo lattice model with correlated
conduction electrons \cite{Schork99}. In a very recent paper this
technique has also been employed to analyze the electronic
structure of the ruthenate alloy series Ca$_{2-x}$Sr$_x$RuO$_4$
\cite{rut}.

All the results presented in the following refer to the case of a
total particle density $n=n_F+2n_B$ equal to 2, with $n_F$ and
$n_B$ being respectively computed from the fermionic and the
bosonic impurity propagators. All energies are measured in units
of the bare bandwidth $2D$. In order to emphasize the effects of
the competition between the boson-fermion coupling and the on-site
Coulomb repulsion, we will mainly consider parameter regimes where
$g$ and $U$ are comparable. We want also to point out that in
these regimes the application of the conventional NCA approach is
expected to give reliable results. This has been checked resorting
to a validity criterion which is discussed at length in
ref.\cite{hallbfm} and which, for brevity, will not be repeated
here. Finally, we have also verified that in the limit of
vanishing boson-fermion coupling, the results for the Hubbard
model obtained in Ref.\cite{Pru93}, where the same DMFT+NCA
algorythm is used, are correctly reproduced.

%%%%%%%%%%%%%%%%%%%%%%%%%%%%%%%%%%%%%%%%%%%%%%%%%%%%%%%%%%%%%%%%%%%%%%
\begin{figure}
\vspace{3.5cm} \centerline{\epsfxsize=8cm \epsfysize=7cm
\epsfbox{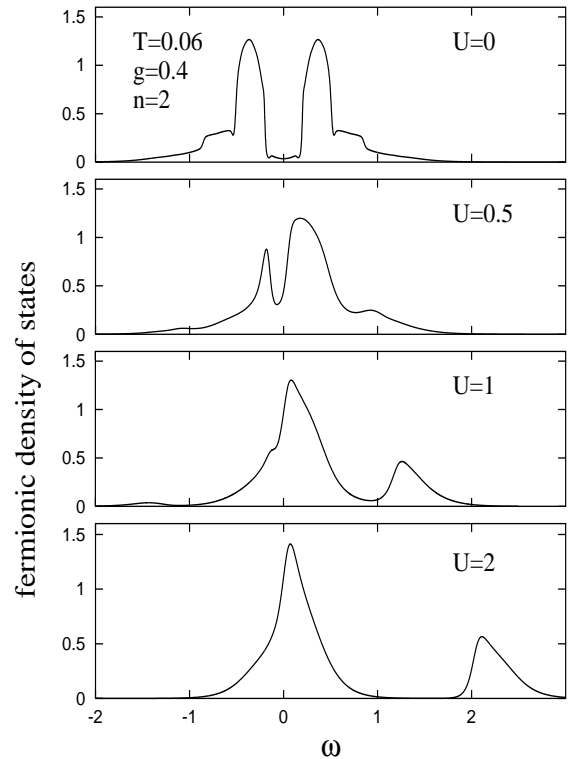}} \caption{Fermionic density of states
for $T=0.06$, $g=0.4$ and several values of $U$. In the case $U=0$
the site energies satisfy the symmetric limit condition
$E_0=\varepsilon_0=0$.} \label{dosu}
\end{figure}
%%%%%%%%%%%%%%%%%%%%%%%%%%%%%%%%%%%%%%%%%%%%%%%%%%%%%%%%%%%%%%%%%%%%%%

We start by discussing how the inclusion of $U$ affects the
pseudogap phase as described by the BFM studied in
Refs.\cite{bfm,pseu,RRR,hallbfm}. In Fig.\ref{dosu} we have
plotted the fermionic density of states for several values of the
on-site repulsion (the energies are measured with respect to the
chemical potential). We start at $U=0$ from the fully symmetric
case $E_0=\varepsilon_0=0$ (top panel), which implies
$n_F=2n_B=1$, and then consider finite values of $U$ while keeping
the bosonic bare energy $\Delta_B$ fixed (the finite-$U$ symmetric
limit conditions $E_0=2\varepsilon_0+U=0$ are consequently not
satisfied). We see that the widely open pseudogap showing up for
$U=0$ is gradually filled as increasing values of $U$ are
considered. At the same time an Hubbard correlation gap of the
order of $U$ develops well beyond the chemical potential, thus
without substantially affecting the low-energy physics. As already
pointed out in the discussion of the atomic limit, this is due to
the fact that the on-site repulsion, tending to inhibit the
boson-fermion exchange mechanism, pushes the system in a regime
where the coupling between the two components is reduced and, for
the parameter choice considered here, the almost pure bosonic
level remains pinned below the fermionic ones (these latter being
splitted by $U$). Starting from the $U=0$ fully symmetric case
considered in the upper panel, this also implies that much of the
fermionic spectral weight is transferred at higher and higher
energies as $U$ is increased, this leading to a substantial
reduction of the fermionic occupation with respect to the bosonic
one. We thus see that for the system is energetically more
favourable to move away from the ``insulating'' half-filled case,
creating holes in the fermionic configuration and thus allowing a
smooth transition towards a metallic behavior.

%%%%%%%%%%%%%%%%%%%%%%%%%%%%%%%%%%%%%%%%%%%%%%%%%%%%%%%%%%%%%%%%%%%%%%
\begin{figure}
\centerline{\epsfxsize=7cm  \epsfbox{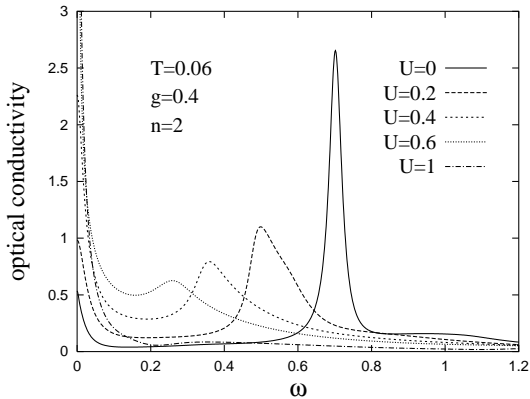}} \caption{Real
part of the optical conductivity for the same parameters as in
Fig.\ref{dosu} and several values of $U$.} \label{optu}
\end{figure}
%%%%%%%%%%%%%%%%%%%%%%%%%%%%%%%%%%%%%%%%%%%%%%%%%%%%%%%%%%%%%%%%%%%%%%

These considerations find further support in the behavior of the
real part of the optical conductivity, which in the limit of
infinite dimensions becomes directly connected to the fermionic
single-particle spectral function $A_F(\varepsilon_k,\omega)$
according to the relation
\begin{eqnarray}
\label{opt} \sigma(\omega)  & = & \pi \int\! d\varepsilon
\,\rho(\varepsilon)\int\! d\omega ' \,
A_F(\varepsilon,\omega')A_F(\varepsilon,\omega+\omega') \,\times
\nonumber \\ & & \times \; {1\over\omega}
\left[n_F(\omega')-n_F(\omega+\omega')\right]
\end{eqnarray}
(here the sum over momenta has again been expressed as an energy
integration over $\rho(\varepsilon)$). From the results reported
in Fig.\ref{optu} we see that a situation typical of that expected
for an insulator is found for $U\to 0$, that is, we obtain a small
d.c. conductivity $\sigma(0)$ with a peak at $\omega\simeq 2g$
reflecting transitions between the bonding and the antibonding
resonances exhibited by the density of states (see
Fig.\ref{dosu}). This peak gradually disappears as $U$ is
increased, with a concomitant Drude-like accumulation of weight al
low $\omega$, clearly associated with the establishment of
metallic properties.

%%%%%%%%%%%%%%%%%%%%%%%%%%%%%%%%%%%%%%%%%%%%%%%%%%%%%%%%%%%%%%%%%%%%%%
\begin{figure}
\vspace{5cm} \centerline{\epsfxsize=8cm \epsfbox{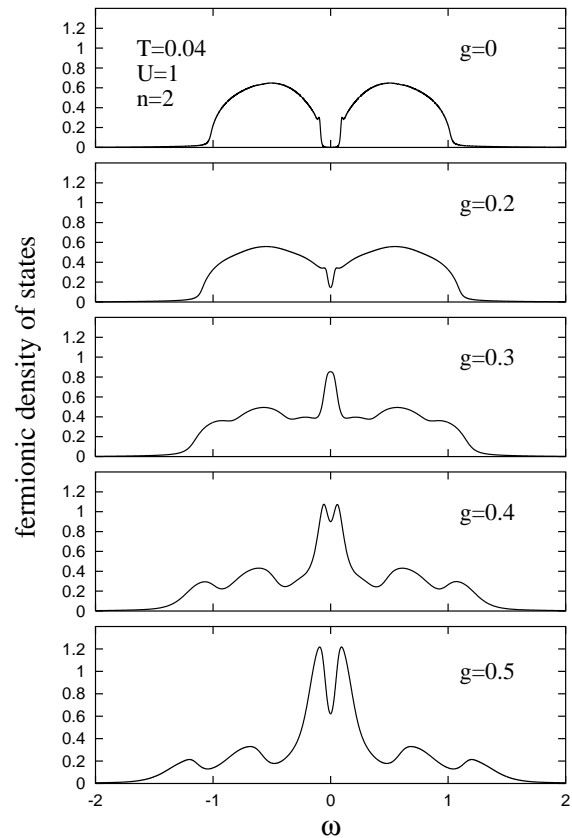}}
\caption{Fermionic density of states in the symmetric limit
($E_0=2\varepsilon_0+U=0$) for $T=0.04$, $U=1$ and several values
of $g$.} \label{dosg}
\end{figure}
%%%%%%%%%%%%%%%%%%%%%%%%%%%%%%%%%%%%%%%%%%%%%%%%%%%%%%%%%%%%%%%%%%%%%%

\begin{figure}
\vspace{5cm} \centerline{\epsfxsize=8cm \epsfbox{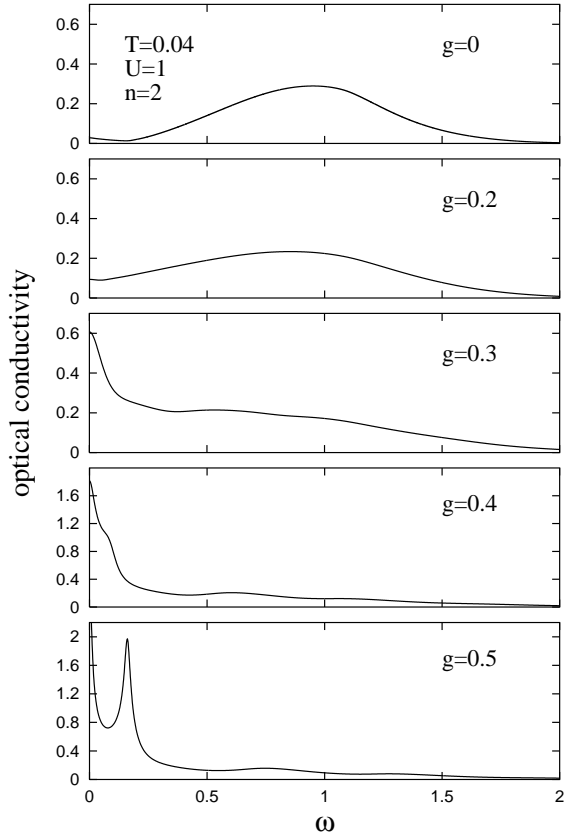}}
\caption{Real part of the optical conductivity for the same
parameters as in Fig.\ref{dosg}.} \label{optg}
\end{figure}

We have seen so far that if one starts from the symmetric limit of
the BFM, characterized by a charge-exchange induced pseudogap in
the excitation spectrum, then the introduction of the on-site
Coulomb repulsion between fermions leads to a transition from an
insulating state to a metallic state with lower carrier density.
Recalling that in the limit $g=0$ the fermionic part of the model
defined by the Hamiltonian (\ref{ham}) becomes equal to the
Hubbard model, we now reverse the point of view, analyzing how the
physics of the Hubbard model in the half-filled insulating phase
is modified by the inclusion of the boson-fermion coupling. To
this purpose, we assume to start from a parameter regime where, in
the absence of $g$, the correlation gap is fully openin a
half-filled configuration. Then we turn on the effect of $g$ under
the assumption that the bosonic energy level is superimposed just
at the middle of the unperturbed fermionic band, in such a way to
ensure a particle density equally distributed among fermions and
bosons ($n_F=2n_B=1$). As previously mentioned, this is realized
when the symmetric limit conditions $E_0=0$ and
$2\varepsilon_0+U=0$ are simultaneously satisfied. The
corresponding behavior of the fermionic density of states and of
the real part of the optical conductivity is illustrated in
Figs.\ref{dosg} and \ref{optg}, respectively.

%%%%%%%%%%%%%%%%%%%%%%%%%%%%%%%%%%%%%%%%%%%%%%%%%%%%%%%%%%%%%%%%%%%%%%
%%%%%%%%%%%%%%%%%%%%%%%%%%%%%%%%%%%%%%%%%%%%%%%%%%%%%%%%%%%%%%%%%%%%%%

\begin{figure}
%\vspace{5cm}
\centerline{\epsfxsize=6cm \epsfbox{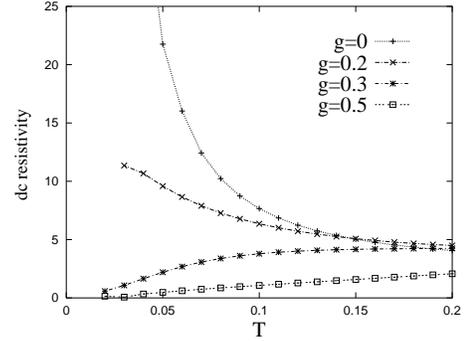}}
\caption{Temperature dependence of the dc resistivity in the
symmetric limit for $U=1$ and several values of $g$.} \label{dct}
\end{figure}

\noindent We see that, as $g$ is increased from zero, the Hubbard
gap is gradually closed as a consequence of the fact that the
boson-fermion coupling tends to favor configurations with double
fermionic occupations. As a result, this mechanism tends to create
empty sites in the half-filled configuration, thus promoting the
fermionic itinerancy. For suitably high values of $g$ a narrow
pseudogap at the chemical potential tends to form again, driving
back the system in a state where the metallic behavior is
gradually weakened as $T$ is decreased. The substantial change in
the transport properties induced by the boson-fermion coupling is
also evident from the temperature behavior of the dc resistivity,
plotted in Fig.\ref{dct} (due to numerical difficulties our NCA
algorythm is currently not able to provide stable enough results
at temperatures lower than those considered in the figure).

%%%%%%%%%%%%%%%%%%%%%%%%%%%%%%%%%%%%%%%%%%%%%%%%%%%%%%%%%%%%%%%%%%%%%%
%%%%%%%%%%%%%%%%%%%%%%%%%%%%%%%%%%%%%%%%%%%%%%%%%%%%%%%%%%%%%%%%%%%%%%

%%%%%%%%%%%%%%%%%%%%%%%%%%%%%%%%%%%%%%%%%%%%%%%%%%%%%%%%%%%%%%%%%%%%%%
\begin{figure}
\vspace{4.5cm} \centerline{\epsfxsize=8cm \epsfbox{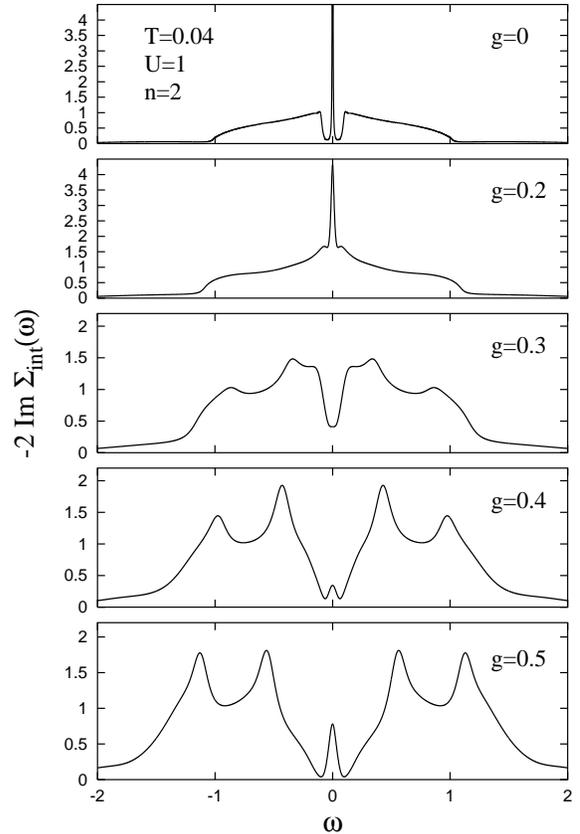}}
\caption{Imaginary part of the fermionic self-energy for the same
parameters as in Fig.\ref{dosg}.} \label{selfg}
\end{figure}
%%%%%%%%%%%%%%%%%%%%%%%%%%%%%%%%%%%%%%%%%%%%%%%%%%%%%%%%%%%%%%%%%%%%%%

\noindent We can see that, while at $g=0$ and for low values of
$g$ the system exhibits the expected semiconductor-like behavior,
with $\sigma(0)^{-1}$ going up as $T$ is decreased, on the other
hand higher values of $g$ give rise to a metallic-like behavior
characterized by a decrease of $\sigma(0)^{-1}$ as $T\to 0$.
Nonetheless, an important qualitative difference should be noted
between the cases of intermediate and high values of $g$. The
development of a dip at the chemical potential above approximately
$g=0.4$ pushes the system toward a state which, though still
showing metallic properties, is characterized by strong deviations
from the Fermi liquid behavior. This is particularly evident when
plotting the imaginary part of the fermionic self-energy for the
same parameter choices as in Fig.\ref{dosg}. We can see from
Fig.\ref{selfg} that $-{\rm Im}\,\Sigma_{int}(\omega)$ exhibits
for $g=0.3$ a parabolic minimum at $\mu$ (though with ${\rm
Im}\,\Sigma(0)_{int}\ne 0$) which is the manifestation of a regime
close to that of a normal Fermi liquid. When higher values of $g$
are considered, the behavior becomes significantly different, with
the minimum being replaced by a narrow peak associated with a
resonant scattering of the fermionic quasiparticles with the
bosons.

In conclusion, we have analyzed a two-component model where
correlated fermions and localized hard-core bosons interact
through a mechanism by which bosons decay in two fermions with
opposite spins and vice-versa. A special attention has been
devoted to how the interplay between the boson-fermion coupling
and the on-site Coulomb repulsion between fermions affects the
spectral and the transport properties. Interesting crossovers from
a pseudogap phase to a metallic regime, and vice-versa, have been
found, as driven by the competition between the above mentioned
local interactions. We emphasize that the way we have taken into
account the effect of the on-site Coulomb interaction among
fermions is not realistic if the model is applied to the
description of high-$T_c$ superconductors. A more proper way would
rather consist in defining the model on a lattice with two kinds
of sites, one associated to the dielectric layers and the other
one to the copper-oxygen planes. The Coulomb repulsion should then
be introduced to describe correlation effects among holes in the
latter, with the charge exchange mechanism now involving two
strongly correlated fermions sitting on neighboring sites, rather
than on the same site. A study along these lines is planned for
the near future.

\bigskip

\noindent{\bf Acknowledgements}

I would like to express my sincere thanks to J. Ranninger for all
the stimulating discussions on the boson-fermion model that we had
in the last years. I would also like to thank M. Cuoco and C. Noce
for helpful discussions and a critical reading of the manuscript.

\end{multicols}


\begin{thebibliography}{99}

\bibitem{Mahan} See for instance: G.D. Mahan, {\it Many Particle
Physics} (Plenum Pubishing, New York, 1981), and references
therein.
\bibitem{Ran85} J. Ranninger and S. Robaszkiewicz, Physica B {\bf
135}, 468 (1985).
\bibitem{bfm} J. Ranninger, J.M. Robin, and H. Eschrig, Phys. Rev.
Lett. {\bf 74}, 4027 (1995).
\bibitem{pseu} J. Ranninger and J.M. Robin, Phys. Rev. B {\bf 53},
R11961 (1996); {ibid.} {\bf 56}, 8330 (1997); P. Devillard and J.
Ranninger, Phys. Rev. Lett. {\bf 84}, 5200 (2000).
\bibitem{RRR} J.M. Robin, A. Romano and J. Ranninger, Phys. Rev.
Lett. {\bf 81}, 2755 (1998).
\bibitem{hallbfm} A. Romano and J. Ranninger, Phys. Rev. B
{\bf 62}, 4066 (2000).
\bibitem{Geo92} A. Georges and G. Kotliar, Phys. Rev. B {\bf 45},
6479 (1992).
\bibitem{Jar92} M. Jarrell, Phys. Rev. Lett. {\bf 69}, 168
(1992).
\bibitem{Jar93} M. Jarrell and Th. Pruschke, Z. Phys. B
{\bf 90}, 187 (1993).
\bibitem{DMFT} A. Georges, G. Kotliar, W. Krauth and M.
Rozenberg, Rev. Mod. Phys. {\bf 68}, 13 (1996).
\bibitem{Geb} F. Gebhard, "The Mott Metal-Insulator Transition"
(Springer Verlag, Berlin, 1997).
\bibitem{NCA} N.E. Bickers, Rev. Mod. Phys. {\bf 59}, 845 (1987);
N.E. Bickers, D.L. Cox and D.L. Wilkins, Phys. Rev. B {\bf 36},
2036 (1987); Th. Pruschke and N. Grewe, Z. Phys. B {\bf 74}, 439
(1989).
\bibitem{elphdmft} J.K. Freericks, M. Jarrell, and D.J. Scalapino,
Phys. Rev. B {\bf 48}, 6302 (1993) and Europhys. Lett. {\bf 25},
37 (1994); J.K. Freericks and M. Jarrell, Phys. Rev. B {\bf 50},
6939 (1994).
\bibitem{Ciuchi} S. Ciuchi, F. De Pasquale, S. Fratini, and
D. Feinberg, Phys. Rev. B {\bf 56}, 4494 (1997).
\bibitem{Millis} A.J. Millis, R. Mueller, and B.I. Shraiman, Phys.
Rev. B {\bf 54}, 5389 (1996).
\bibitem{Kot} Y. Motome and G. Kotliar, cond-mat/0005395.
\bibitem{HH} J. Takimoto and Y. Toyozawa, J. Phys. Soc. Japan {\bf
52}, 4331 (1983); J. Zhong and H.B. Sch\"uttler, Phys. Rev. Lett.
{\bf 69}, 1600 (1992); U. Trapper, H. Fehske, M. Deeg, and H.
B\"{u}ttner, Z. Phys. B {\bf 93}, 465 (1994); G. Wellein, H.
R\"{o}der, and H. Fehske, Phys. Rev. B {\bf 53}, 9666 (1996); J.K.
Freericks amd M. Jarrell, Phys. Rev. Lett. {\bf 75}, 2570 (1995);
T. Hotta and Y. Takada, Physica B {\bf 230-232}, 1037 (1997); A.
La Magna and R. Pucci, Phys. Rev. B {\bf 55}, 14886 (1997); M.
Capone, M. Grilli, and W. Stephan, cond-mat/9902317; J.
Bon{\v{c}}a, T. Katra{\v{s}}nik, and S.A. Trugman, Phys. Rev.
Lett. {\bf 84}, 3153 (2000); M. Acquarone, M. Cuoco, C. Noce, and
A. Romano, Phys. Rev. B, 15 Jan. 2001 issue.
\bibitem{Dom98} T. Domanski, J. Ranninger and J.M. Robin, Solid
State Commun. {\bf 105}, 473 (1998).
\bibitem{DCA} M.H. Hettler, A.N. Tahvildarzadeh, M. Jarrell,
T. Pruschke, H.R. Krishnamurthy, Phys. Rev. B {\bf 58}, 7475
(1998); M.H. Hettler, M. Mukherjee, M. Jarrell, and H.R.
Krishnamurthy, Phys. Rev. B {\bf 61}, 12739 (2000); Th. Maier, M.
Jarrell, Th. Pruschke, and J. Keller, Eur. Phys. J. B {\bf 13},
613 (2000); Th. Maier, M. Jarrell, Th. Pruschke, and J. Keller,
Phys. Rev. Lett. {\bf 85}, 1524 (2000).
\bibitem{Jar00} M. Jarrell, Th. Maier, M.H. Hettler, and A.N.
Tahvildarzadeh, cond-mat/0011282.
\bibitem{Metz} W. Metzner amd D. Vollhardt, Phys. Rev. Lett. {\bf
62}, 324 (1989).
\bibitem{Mul89} E. M\"uller-Hartmann, Z. Phys. B {\bf 74}, 507
(1989).
\bibitem{Schork} T. Schork and S. Blawid, Phys. Rev. B {\bf 56},
6559 (1997).
\bibitem{Schork99} T. Schork, S. Blawid, and J. Igarashi,
Phys. Rev. B {\bf 59}, 9888 (1999).
\bibitem{Pru93} Th. Pruschke, D.L. Cox and M. Jarrell, Phys. Rev.
B {\bf 47}, 3553 (1993).
\bibitem{Avignon} P. Lombardo, J. Schmalian, M. Avignon, and K.-H.
Bennemann, Phys. Rev. B {\bf 54}, 5317 (1997).
\bibitem{rut} V.I. Anisimov, I.A. Nekrasov, D.E. Kondakov, T.M.
Rice, and M. Sigrist, cond-mat/0011460.

\end{thebibliography}
\end{document}